\documentclass[10pt, conference]{IEEEtran}
\IEEEoverridecommandlockouts

\usepackage{cite}
\usepackage{amsmath,amssymb,amsfonts}
\usepackage{algorithmic}
\usepackage{graphicx}
\usepackage{textcomp}
\usepackage{xcolor}
\usepackage{bm}

\usepackage[T1]{fontenc}
\usepackage[ruled,linesnumbered,noend]{algorithm2e}

\usepackage[bookmarks=true,breaklinks=true,letterpaper=true,colorlinks,linkcolor=black,citecolor=blue,urlcolor=black]{hyperref}

\usepackage{comment}
\newcommand{\circleNumber}[1]{
{\large \textcircled{\small #1}} }

\newcommand{\insertFigure}[2]{
    \begin{figure}[t]
\setlength{\abovecaptionskip}{-3pt}
\setlength{\belowcaptionskip}{-1pt}
        \centering
        \includegraphics[width=\linewidth]{\FIGDIR/#1}
	\vspace{-2mm}
        \caption{#2}
	\vspace{-2mm}
        \label{fig:#1}
    \end{figure}
}

\newcommand{\insertWideFigure}[2]{
    \begin{figure*}
\setlength{\abovecaptionskip}{-3pt}
\setlength{\belowcaptionskip}{-1pt}
        \centering
        \includegraphics[width=\textwidth]{\FIGDIR/#1}
	\vspace{-2mm}
        \caption{#2}
	\vspace{-2mm}
        \label{fig:#1}
    \end{figure*}
}
\newcommand{\tileload}{\textit{tileload}\xspace}
\newcommand{\tilestore}{\textit{tilestore}\xspace}

\newcommand{\rasatl}{\textit{rasa\_tl}\xspace}
\newcommand{\rasats}{\textit{rasa\_ts}\xspace}
\newcommand{\rasamm}{\textit{rasa\_mm}\xspace}

\newcommand{\TODO}[1]{\textcolor{red}{TODO: #1}}
\newcommand{\GJ}[1]{\textcolor{magenta}{GJ: #1}}
\newcommand{\EQ}[1]{\textcolor{blue}{EQ: #1}}
\newcommand{\AS}[1]{\textcolor{orange}{AS: #1}}

\newcommand{\HK}[1]{\textcolor{brown}{HK: #1}}
\newcommand{\TK}[1]{\textcolor{red}{TK: #1}}
\newcommand{\sree}[1]{\textcolor{purple}{SS: #1}}

\newcommand{\RASA}{\textsc{RASA}\xspace}
\newcommand{\OptOne}{\textsc{PIPE}\xspace}
\newcommand{\OptTwo}{\textsc{WLBP}\xspace}
\newcommand{\OptThree}{\textsc{WLS}\xspace}

\def\FIGDIR{./figures}

\begin{document}

\title{RASA: Efficient Register-Aware Systolic Array Matrix Engine for CPU\\
\thanks{This work was supported by an award from Intel. We thank the LIBXSMM team (Alexander Heinecke and Hans Pabst) for their help with our methodology.}
}

\author{\IEEEauthorblockN{Geonhwa Jeong\IEEEauthorrefmark{1}, Eric Qin\IEEEauthorrefmark{1}, Ananda Samajdar\IEEEauthorrefmark{1}, Christopher J. Hughes\IEEEauthorrefmark{2}, \\
Sreenivas Subramoney\IEEEauthorrefmark{2}, Hyesoon Kim\IEEEauthorrefmark{1} and Tushar Krishna\IEEEauthorrefmark{1} 
    \IEEEauthorblockA{\IEEEauthorrefmark{1}\textit{Georgia Institute of Technology}, Atlanta, USA \\
    \{geonhwa.jeong, ecqin, anandsamajdar\}@gatech.edu,
    hyesoon@cc.gatech.edu, tushar@ece.gatech.edu}
    \IEEEauthorblockA{\IEEEauthorrefmark{2}\textit{Intel Labs}, 
    \{christopher.j.hughes, sreenivas.subramoney\}@intel.com}
    }
    }
\vspace{-2mm}

\maketitle

\begin{abstract}
As AI-based applications become pervasive, CPU vendors are starting to incorporate matrix engines within the datapath to boost efficiency. Systolic arrays have been the premier architectural choice as matrix engines in offload accelerators. However, we demonstrate that incorporating them inside CPUs can introduce under-utilization and stalls due to limited register storage to amortize the fill and drain times of the array. To address this, we propose \RASA, Register-Aware Systolic Array. We develop techniques to divide an execution stage into several sub-stages and overlap instructions to hide overheads and run them concurrently. \RASA-based designs improve performance significantly with negligible area and power overhead.

\end{abstract}

\section{Introduction}

In diverse areas including, but not limited to, computer vision, natural language processing, and personal recommendation, Deep Learning (DL) models have shown dramatic performance, even exceeding that of humans for some tasks. 
DL workloads are both notoriously compute hungry and commonplace in applications, motivating enhancements to hardware and software platforms to improve performance and energy efficiency~\cite{jouppi2017datacenter, eyeriss_isca, kwon2018maeri}.
%
General Matrix-Matrix Multiplication (GEMM), a staple of high-performance computing, is also a critical building block for many DL applications including Transformers for natural language processing, Multi-Layer Perceptrons (MLPs) for recommendation models, and Convolutional Neural Networks (CNNs) for computer vision tasks.
%
For DL workloads, GEMM performance and energy-efficiency are sufficiently important that special-purpose hardware support has become common.  
A systolic array is one of the most efficient structures for dense GEMMs given its simple construction, high concurrency, and ability to efficiently exploit the inherent data reuse in the computation. 
To that end, most DL accelerators include systolic arrays for GEMM, including
Google's TPU, Xilinx's XDNN, and Habana's Goya.
Further, even less specialized platforms, such as Nvidia GPUs, also use (small) systolic arrays~\cite{tensorcore}.

%
In this work, we explore adding a systolic array for GEMMs to a CPU for DL workloads.
The decision to go with systolic arrays is motivated by simple construction and small control overhead while simultaneously achieving high throughput compared to other accelerators \cite{jouppi2017datacenter}. 
While custom DL accelerators' and GPUs' are highly popular for running compute-intensive DL workloads in the datacenter and edge due to their high compute density 
and memory bandwidth, 
studies have shown that most 
DL workloads are actually running on CPUs today~\cite{wu2019}, given their pervasiveness.
CPUs trade off reduced compute density and energy efficiency for improved generality and programmability.
%
%
To meet the huge demand for running DNNs efficiently, CPU vendors like IBM, ARM, and Intel have begun introducing DNN-specific optimizations.
Recently, Intel announced the inclusion of Advanced Matrix Extensions (AMX)~\cite{intel20isa} to their ISA, to allow software for CPUs to express deep learning GEMMs with a greatly reduced instruction count.
Despite this intense commercial activity, to the best of our knowledge, no previous work explores the design trade-offs of integrating
systolic arrays in CPUs for GEMMs. 
Unlike standalone accelerator hardwares, a systolic array housed within the CPU core is subject to constraints of the existing conventional memory system. 
We identify that 
the limited size of registers inside CPUs to feed a systolic array can lead to increased under-utilization of the array during 
its fill and drain times, and even during some of output generation times. To address this, we
propose a systolic array based functional unit called \textit{\underline{R}egister-\underline{A}ware \underline{S}ystolic \underline{A}rray Matrix Engine} (\RASA) and develop techniques to divide an execution stage into several sub-stages for properly overlapping instructions.

In summary, the key contributions of our work are as follows.
\textbf{(i)} \textit{We first introduce an efficient register-aware systolic array based matrix engine for CPU and show how it can be driven by simple GEMM instructions.}
\textbf{(ii)} \textit{We identify challenges for integrating a systolic array as a matrix engine in a CPU and propose a set of \RASA-Control and \RASA-Data optimizations.
\RASA-Control extracts higher performance by introducing novel pipelining and bypassing schemes on the conventional systolic array, whereas \RASA-Data optimizations include microarchitectural changes in processing elements.}
\textbf{(iii)} \textit{We implement \RASA~based designs on RTL with Nangate-15nm library and evaluate with various real DL workloads.}
Our results show that \RASA-Control optimizations reduce runtime by 30.9\% with control hardware changes only, and using both \RASA-Control and \RASA-Data optimizations improves 79.2\% in runtime while consuming a total 0.847$mm^2$ in area.

\section{Background}


\subsection{Deep Neural Networks (DNNs) and GEMM}

DNNs are comprised of a series of layers, where each layer represents a particular computation. 
The most common computationally intensive layers are 
fully connected (FC) and convolutional layers.
Multi-Layer Perceptrons (MLP) are composed of a number of FC layers, and are at the heart of many modern DNNs, including recommendation models such as DLRM ~\cite{maxim19dlrm} 
and natural language processing models like BERT ~\cite{jacob2019bert}. 
If we group multiple inputs (i.e., {\em batch}) through an FC layer, the computation becomes (primarily) a GEMM. 
Convolutional Neural Networks (CNNs) are dominated by convolutional layers and are extremely popular in image and video processing. 
Many implementations ``lower'' the convolution computation to GEMMs \cite{Evangelos20ipdps}.



\subsection{SIMD/Matrix Extensions in CPUs for running GEMM}
Popular high-performance GEMM implementations heavily leverage Single Instruction Multiple Data (SIMD), especially fused multiply-add (FMA) operations. 
Due to the performance and efficiency demands of DL workloads, CPU vendors have begun introducing enhancements to their SIMD hardware for GEMMs, including Intel's AVX512\_VNNI instructions~\cite{intel2018vnni}, IBM Power's outer product instructions, and support for smaller data types such as BF16. 

Recently, CPU vendors have started introducing support in the ISA for running GEMM. For the purpose of illustration, we use Intel's Advanced Matrix Extensions (AMX)~\cite{intel20isa} as a reference design, though 
our analysis and conclusions are more broadly applicable.
AMX includes eight 1KB 2D registers (called ``tile registers") 
and instructions to operate on them~\cite{intel20isa}.  For software operating on a 2D array, i.e., a matrix, each tile register may hold an entire small matrix, or, for larger matrices, a sub-matrix (tile or block).
Tile register can be moved from/to memory with the \tileload and \tilestore instructions, respectively---a tile in memory is a set of up to 16 chunks of data up to 64B each, separated by a fixed stride.  
Given this ISA support for GEMM,
the goal of this work is to study microarchitectural implementations
for GEMM engines
to natively run GEMMs in CPUs.

\subsection{Baseline Matrix Engine: Systolic Array}
\label{sec:systolic_background}
Systolic arrays are two-dimensional arrays of processing elements (PEs), connected with peer-to-peer links. 
They provide high concurrency and high compute to memory ratios due to their regular structure and simple construction. This enables attaining high performance while achieving favorable energy efficiencies. Therefore, several industrial and academic accelerators use them~\cite{jouppi2017datacenter,bahar20iccd,scalesim-ispass}.

A GEMM kernel multiplies a $M \times K$ matrix $\bm{A}$ with a $K \times N$ matrix $\bm{B}$ to generate the output $M \times N$ matrix $\bm{C}$.
A mapping determines how the operands are fed into and stored in the array and leads to different {\em dataflows}. As described in a recent work~\cite{scalesim-ispass}, there are three popular classes of dataflows for GEMM, Input Stationary (IS), Weight Stationary (WS), and Output Stationary (OS). 
Each dataflow maps one of $\bm{A}$, $\bm{B}$, and $\bm{C}$, to the 2D array of PEs\footnote{We assume DNN weights are in $\bm{B}$, but software could place them in $A$.}, holding a single element of that matrix in place at a PE throughout execution; elements of the other matrices flow through PEs.
The choice of dataflow affects performance and energy efficiency, with the best option depending on the dimensions of the operands and the parameters of the systolic array~\cite{scalesim-ispass}.
In modern accelerators, WS is generally preferred since it exploits high spatio-temporal reuse of weights \cite{jouppi2017datacenter}. 
As mentioned earlier, when mapping into the array the entire computation will often not be mapped at once but will be mapped in portions or ``tiles''. 
We use the $T_M$, $T_N$ and $T_K$ for the dimensions of the tiles when mapped onto the array.
Each tile will then be mapped onto the in various passes (which we call a ``fold") until all outputs are generated.
In a traditional systolic array, each fold requires several steps such as filling the stationary operands, streaming the non-stationary operands, reducing across MAC units, and draining the generated outputs.
As shown in Fig.~\ref{fig:pe_util_toy.pdf}, for a tiled GEMM on a WS systolic array with $T_K \times T_N$ PEs, the bottom right PE is in the critical path.
It takes $T_K$ cycles to load the stationary weight elements. 
Then, it takes $T_N - 1$ cycles to fill the pipeline and the first input operand to arrive at this PE, and the next $T_M$ cycles are spent on MAC computations for all the incoming input elements. Finally, additional $T_K - 1$ cycles are needed for reduction and ejection of the last output element. The total latency can thus be calculated as~\cite{scalesim-ispass, bahar20iccd}:
\begin{equation} 
Latency_{tot} = 2T_K + T_N + T_M - 2 \label{ws_latency}
\end{equation}

\section{Challenges for Systolic Arrays in CPUs}
\label{Motivation-challenge}
\insertFigure{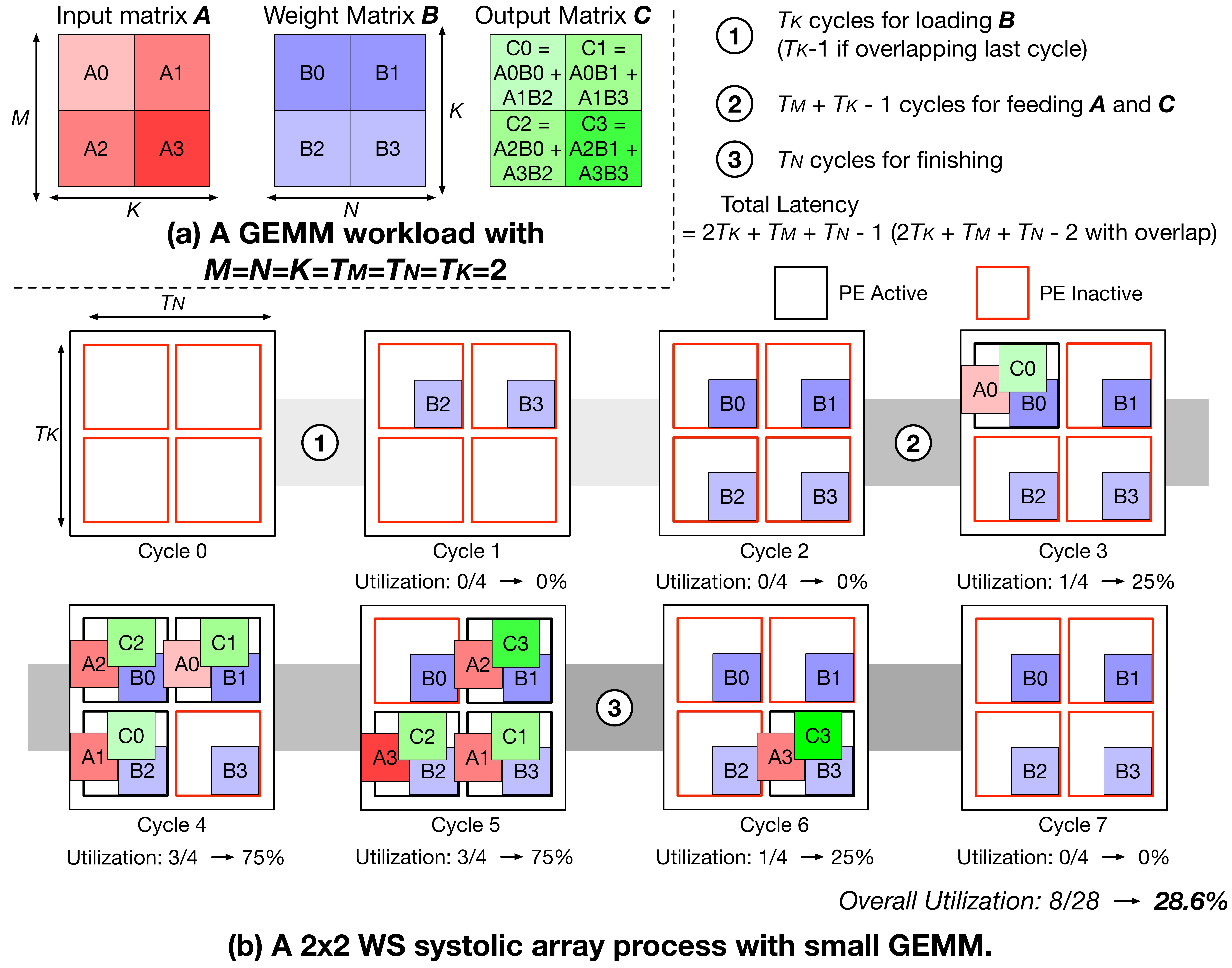}{A 2$\times$2 WS systolic array for processing 2$\times$2 input matrix and weight matrix.}
\insertFigure{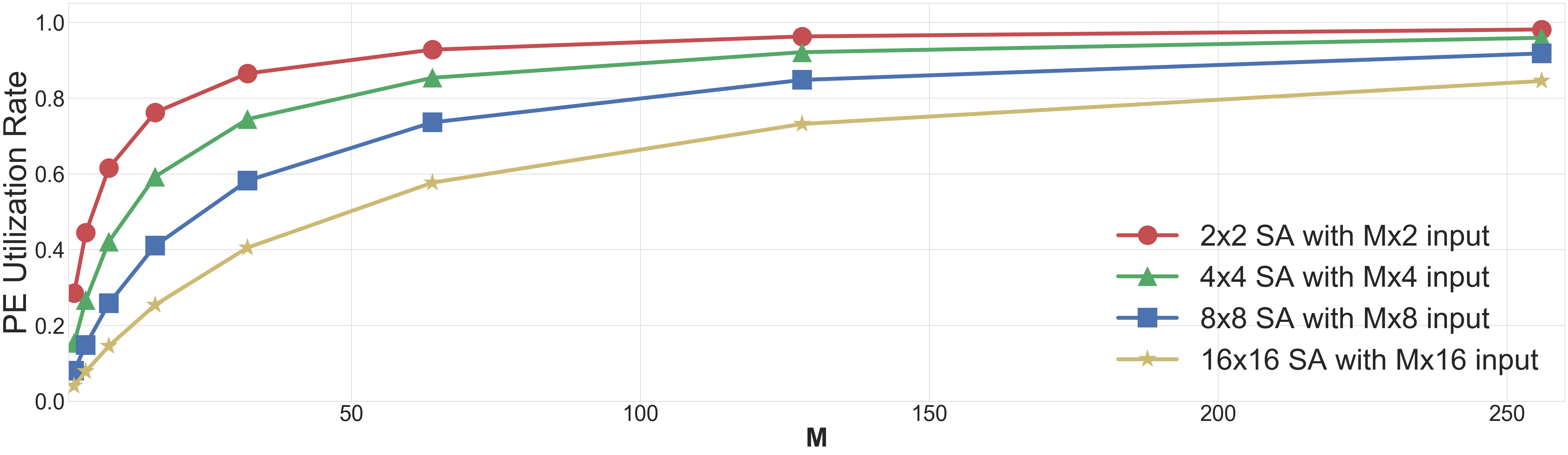}{PE utilization ratio with different input size and SA (Systolic Array) dimensions.}
Systolic arrays, given their two-dimensional construction with neighbor-to-neighbor links, are simple to construct, control, and are capable of high throughput and operand reuse when fully utilized.
However, simple construction leads to challenges with utilization.

\textbf{Under-utilization in Systolic Arrays.}
In Fig.~\ref{fig:pe_util_toy.pdf}, we show a toy example of a 2$\times$2 weight stationary systolic array processing 2$\times$2 input matrices and observe that the average PE utilization is only 28.6\%. 
Though this is a toy example, the challenge of under-utilization in systolic arrays is quite fundamental~\cite{scalesim-ispass}.
There are three cases in which the entire array may not be utilized for all cycles.
(i) \textit{Mapping inefficiency}. Due to the mismatch between the workload and the array dimensions, some compute units are left idle.
(ii) \textit{Memory Stalls}. This occurs when the memory system cannot supply data fast enough to keep up the throughput.
(iii) \textit{Pipeline fill and drain delay}. Recall from \autoref{sec:systolic_background}, 
each PE computes for $T_M$ cycles, 
the rest of the time goes in
fill and drain.
Thus, even if we have 100\% mapping and sufficient memory bandwidth, each MAC is still inactive for the following cycles,
\begin{equation} 
Time_{inactive} = Latency_{tot} - T_M
\label{underutilized_latency}
\end{equation}
Note that all the PEs are idle at the different cycles due to the pipelined movement of operands through the array as shown in Fig.~\ref{fig:pe_util_toy.pdf}.
For cases when $T_M$ is comparatively small, this leads to severe performance degradation.
In Fig.~\ref{fig:pe_util_toy.pdf}(b), each PE is active for $T_M = 2$ cycles and inactive for the remaining $5$ cycles (71\% performance degradation).

\textbf{Combating Under-utilization.}
The focus of this work is on combating under-utilization due to pipeline fill and drain.
Fundamentally, there can be two ways of addressing this.
%
The first is to speed up fill and drain itself 
via richer interconnects like buses~\cite{eyeriss_isca} for single-cycle operand broadcast and 
trees for reduction in logarithmic steps~\cite{kwon2018maeri}.
Unfortunately, this comes at the high area and power overhead~\cite{kwon2018maeri} 
which is not acceptable
within a CPU since it 
has tight area and power constraints.
The second is to use a large tile size for $M$.
As can be seen from \autoref{underutilized_latency}, 
a large $T_M$ can 
make the inactive time percentage of each PE ($1 - \frac{T_M}{Latency_{tot}}$) converge to 0.
The basic idea is once the array is filled up with $\bm{B}$, we keep streaming elements of $\bm{A}$ and $\bm{C}$ through the array, keeping it at peak utilization for a longer time. Fig.~\ref{fig:pe_util_ratio.png} quantifies this effect, showing that under-utilization of the array is alleviated with a large $T_M$. Keeping in mind that the peak throughput of the array is $T_K \times T_N$ operations per cycle, we would like to grow these; Fig.~\ref{fig:pe_util_ratio.png} shows that a large $T_M$ helps even as $T_K$ and $T_N$ grow.



Recent DNN accelerators~\cite{jouppi2017datacenter} have followed the simple strategy of deploying an array with many PEs, i.e., large $T_K$ and $T_N$. To achieve good efficiency, they leverage a very large $T_M$.  For example, the first layer of ResNet50~\cite{kaiming2016resnet}, can be transformed to a GEMM with $M$=47524, $N$=64, $K$=147, allowing even a large systolic array to have reasonable efficiency by employing large tile sizes for $M$.

This cannot be applied on a CPU systolic array, however, since the tile sizes are limited by the size of the tile registers determined by the ISA (e.g., 1KB in Intel AMX). Increasing the size of the tile registers comes with overhead in area and power.
This is a challenge, especially for CPUs due to its nature of general-purpose. 
Area and power devoted to specialized features under the same budget mean lower performance and efficiency for workloads that cannot use them.
Another solution is to feed the array directly from memory bypassing the tile registers completely. But this introduces many challenges for a CPU, including that a CPU must be responsive to interrupts and exceptions.

The limitations on a large $T_M$ for CPUs motivate 
our solution that enables higher utilization despite limitations in register size.

\section{\RASA}
\subsection{System Overview}
\insertFigure{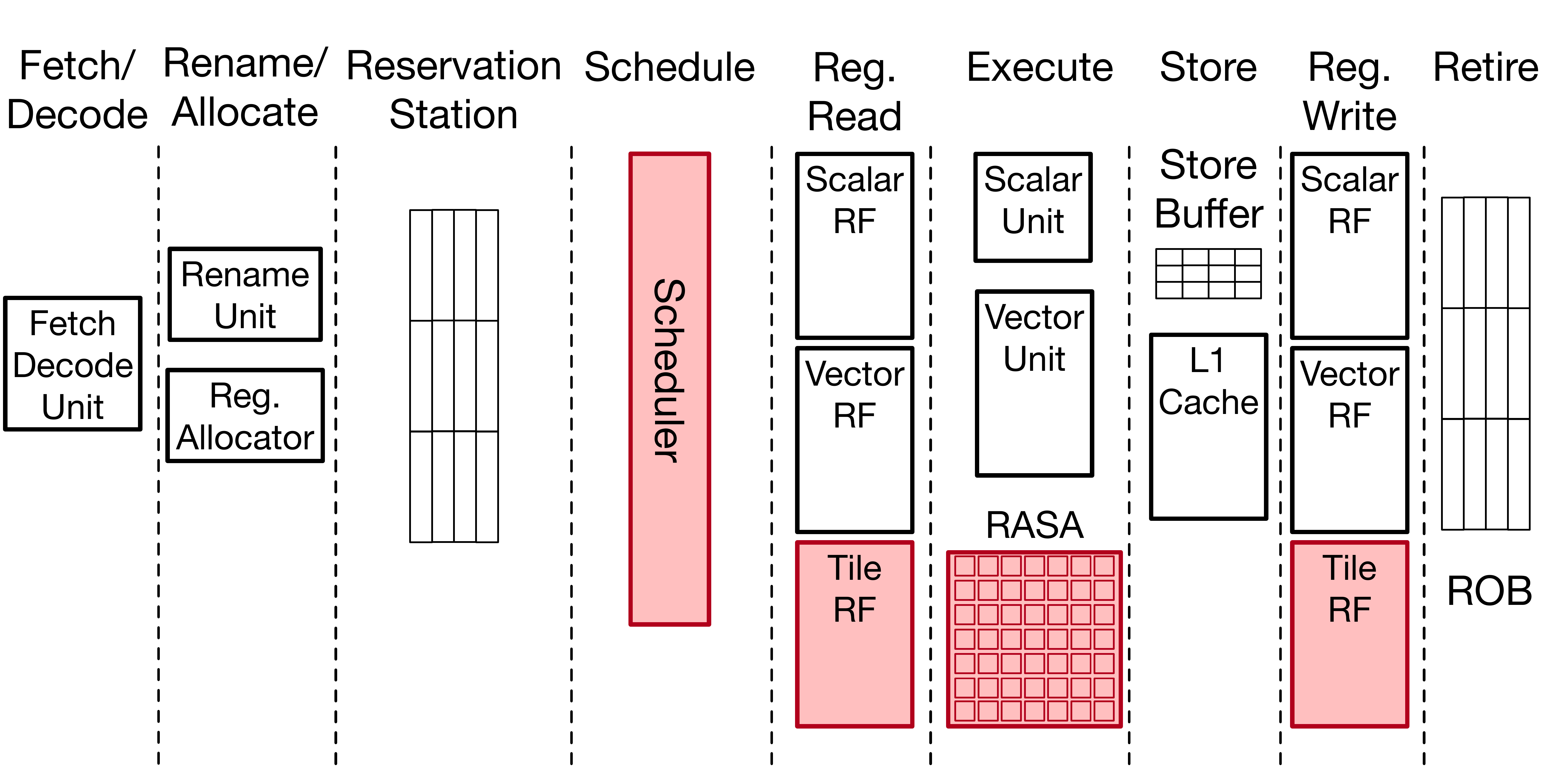}{System overview. Blocks in red include our contributions.}
\begin{algorithm}[t]
\footnotesize
\SetInd{0.45em}{0.45em}
\SetAlgoLined
\DontPrintSemicolon
\KwIn{Tile addresses for 32$\times$32 matrices $\bm{A}, \bm{B}, \bm{C}$, ATile0-1, BTile0-1, CTile0-3 for $\bm{C}$ += $\bm{A}\times\bm{B}$ with BF16 and FP32 for input and output data types.
\\
}


// Step 1. Load C Tiles to tile registers \\
\textit{rasa\_tl} treg0, \textbf{ptr}[CTile0]  \\
\textit{rasa\_tl} treg1, \textbf{ptr}[CTile1] \\
\textit{rasa\_tl} treg2, \textbf{ptr}[CTile2] \\
\textit{rasa\_tl} treg3, \textbf{ptr}[CTile3] \\
// Step 2. Compute partial sums\\
\textit{rasa\_tl} treg4, \textbf{ptr}[BTile0] \\
\textit{rasa\_tl} treg6, \textbf{ptr}[ATile0] \\
\textit{rasa\_mm} treg0, treg6, treg4 \\
\textit{rasa\_tl} treg7, \textbf{ptr}[ATile1] \\
\textit{rasa\_mm} treg1, treg7, treg4 \\
\textit{rasa\_tl} treg5, \textbf{ptr}[BTile1] \\
\textit{rasa\_mm} treg2, treg6, treg5 \\
\textit{rasa\_mm} treg3, treg7, treg5 \\
// Step 3. Store C Tiles back to the memory \\
\textit{rasa\_ts} \textbf{ptr}[CTile0], treg0 \\ 
\textit{rasa\_ts} \textbf{ptr}[CTile1], treg1 \\
\textit{rasa\_ts} \textbf{ptr}[CTile2], treg2 \\
\textit{rasa\_ts} \textbf{ptr}[CTile3], treg3 \\


\caption{A code example with \textsc{RASA}\xspace instructions}
\label{alg:gemm_example}
\end{algorithm} \label{gemm-amx}
\insertWideFigure{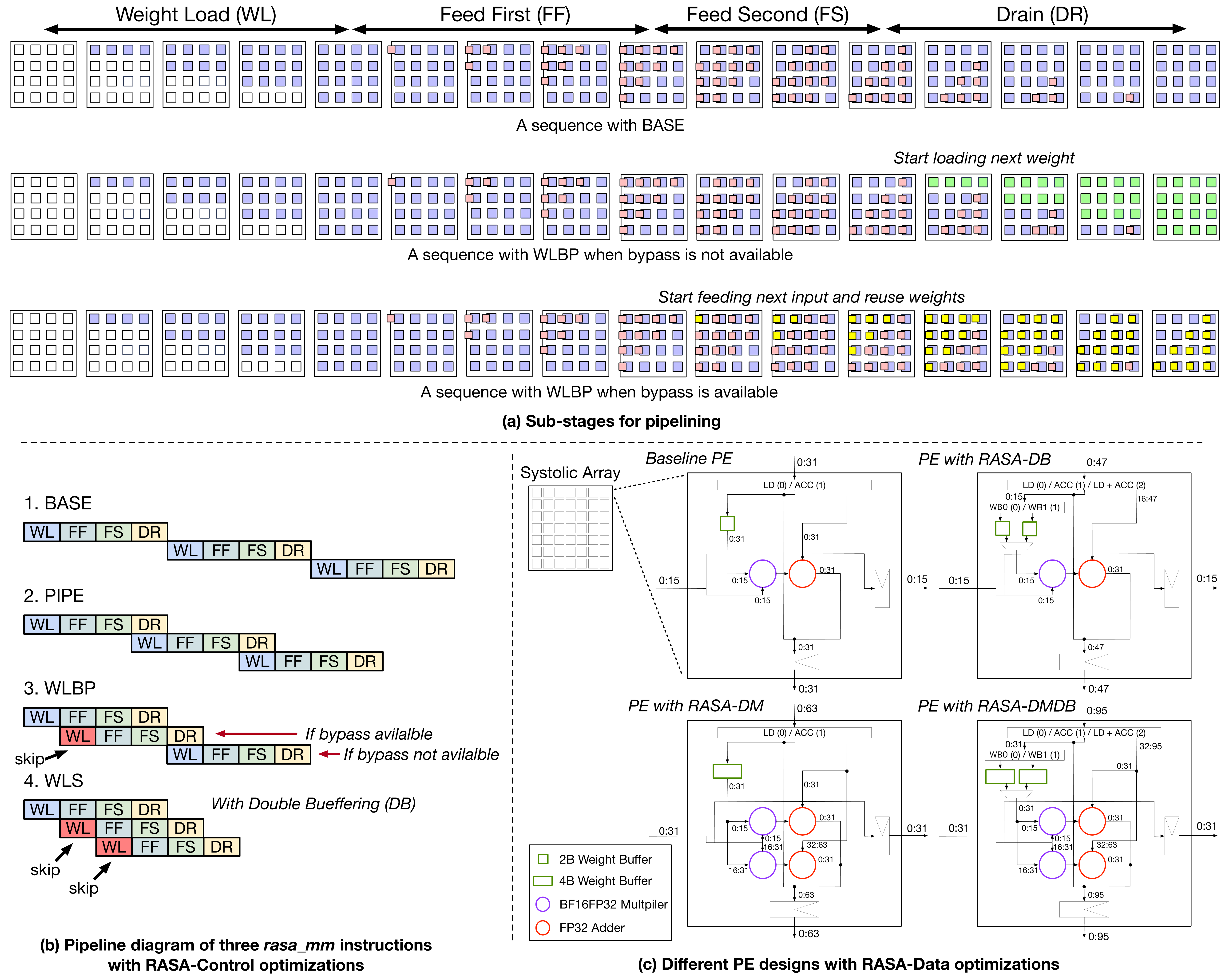}{\RASA overview. (a) shows sub-stages used in \RASA for pipelining. (b) shows three RASA-Control optimizations. We add one-cycle bubble at the end of FS stage to make all stages have same latency for this example for clear explanation. (c) shows three RASA-Data optimizations. 
}
Inspired by Intel AMX, we assume a system with eight architectural tile registers treg0-7, each composed of 16 rows of 64B. We use \rasatl, \rasats, \rasamm as the interface with the matrix engine. \rasatl loads data from memory to the specified register while \rasats stores it back to the memory. 
A matrix multiplication and accumulation can be done by using the \rasamm instruction on tile registers. A high-level system overview is shown in Fig.~\ref{fig:system-overview-alt.pdf}. A simple example code using \RASA instructions is shown in Algorithm~\ref{alg:gemm_example}. For the baseline, we use a WS systolic array with 32$\times$16 PEs and each PE have a single multiply-accumulate (MAC) unit. Fig.~\ref{fig:pe_util_toy.pdf} shows how a WS systolic array works.
All of the operands are read/written from/to tile registers. The hardware first reads $\bm{B}$ a row at a time from the bottom up, inserting data into the north port of the array. 
Next, it inserts values from $\bm{A}$ and $\bm{C}$ into the west and north ports of the systolic array, respectively, in a skewed manner. 
Each cycle, every PE forwards the current $\bm{A}$ element to the east and the previous cycle's result, a partial sum, to the south. The PEs at the bottom of the systolic array produce the final outputs which are written back to the output register.

\subsection{Design}
Fig.~\ref{fig:RASA-overview.pdf}(a) illustrates which inputs and weights are mapped on PEs in each time step when executing two \rasamm instructions in a 4$\times$4 systolic array. PEs become purple when the weights of the first \rasamm instruction is loaded while they change to green when the weights for the second instruction are loaded. Red and yellow boxes correspond to the inputs of the first and second instructions respectively.
We first divide the execution stage of the WS dataflow on a systolic array into four sub-stages and explain how to overlap sub-stages, including which resources should be added.

\circleNumber{1}Weight Load (WL): WS dataflow needs weight values ($\bm{B}$) to be loaded in PEs. Without dedicated links for this, it takes $T_K$ cycles from the top edge of the systolic array to the bottom edge. (The last cycle of WL can be overlapped with the first cycle of FF. In that case, WL effectively takes $T_K-1$ cycles.).
\circleNumber{2}Feed First (FF): 
During this stage, $\bm{A}$ and $\bm{C}$ elements are fed into the systolic array from the left and top. FF ends when we finish feeding the first row of the systolic array. We split the Feed stage here for pipelining.
This stage takes $T_M$ cycles.
\circleNumber{3}Feed Second (FS): 
In FS stage, we finish feeding inputs for the remaining rows of the array. Note that some PEs are now idle, and we will leverage those to start executing the next instruction.
This stage takes $T_K-1$ cycles. 
\circleNumber{4}Drain (DR): In this stage, we let the data in the systolic array finish propagating east and south. Draining the remaining outputs takes $T_N$ cycles.
Without any optimizations or pipelining, \rasamm instructions are completely serialized (as shown in Fig.~\ref{fig:RASA-overview.pdf}(b) BASE), and throughput will be one instruction every $Latency_{tot}$ cycles (Eq.~\ref{ws_latency}).  We call this as the BASE design. 

\textbf{\RASA-Control Optimizations.}
We propose three control optimizations, called \RASA-control, for pipelining multiple \rasamm instructions within a systolic array concurrently as shown in Fig.~\ref{fig:RASA-overview.pdf}(a) and (b).

(i) \textit{Basic Pipelining}  (\OptOne): We observe that the DR stage of previous \rasamm instruction can be overlapped with the WL stage of the next \rasamm instruction. This does not require new hardware in the PEs.

(ii) \textit{Weight Load Bypass} (\OptTwo): We observe that sometimes consecutive \rasamm instructions reuse the $\bm{B}$ (weight) register. For example, in Algorithm 1, lines 9 and 11 reuse treg4 and lines 13 and 14 reuse treg5 as the weight register. If we identify this, we can skip the WL stage in \RASA unless the content in the reused register has been changed in the meantime. 
We add a dirty bit to each tile register to track whether it has been changed after the previous \rasamm. With \OptTwo, if $\bm{B}$ is reused and has a clear dirty bit, we can start the FF stage during the DR stage of the previous instruction. 
We further observe that the FS stage does not fully utilize the left-top PEs, which are needed by the FF stage of the next instruction; thus, when we skip WL, we also allow these stages to be overlapped.

(iii) \textit{Weight Load Skip} (\OptThree):  \OptTwo aggressively pipelines the systolic array, but only when weights are reused.  To achieve this same throughput in other cases, we can do WL for an instruction during the previous instruction's FF; however, this requires extra buffers and links, which will be discussed later. Thus, \OptThree hides WL latency by prefetching.



\textbf{\RASA-Data Optimization.}
In Fig.~\ref{fig:RASA-overview.pdf}(c), we show the baseline PE and the PE designs with RASA-Data optimizations. 
The PEs perform mixed-precision matrix multiplication (BF16 in, FP32 out). The baseline PE design has one multiplier, one adder, and buffers a single weight. \RASA-DB uses PEs with Double Buffering (DB). These use an extra weight buffer and links to enable one of the aforementioned \RASA-Control optimizations, \RASA-WLS.
We also propose RASA-DM, a PE design with a Double Multiplier (DM), and an extra adder. This PE updates two partial sums in parallel; thus, we place a row of adders at the bottom of the systolic array to merge the two partial sums. DM doubles the size of an individual PE, but we compensate by reducing half of the total number of PEs in the DM systolic array implementation.
Finally, \RASA-DMDB includes both DB and DM features. 
We show in Section~\ref{evaluation} that the \RASA-Data optimizations cost negligible area overhead over the baseline.
\section{Evaluation}
\label{evaluation}
To evaluate the proposed design and optimizations, we use Intel AMX instructions as the interface with the proposed design. We use optimized convolutions and MLPs from the LIBXSMM library~\cite{Evangelos20ipdps}, using AVX and AMX. We collected traces with Intel SDE. We implement the \RASA optimizations in MacSim \cite{hyesoon12macsim}, a trace-driven cycle-level simulator with the following configuration: CPU (and NoC) at 2GHz, 16 pipeline stages, ROB size of 97, fetch/issue/retire width of 4, similar to Intel's Skylake.
To focus on the systolic array tradeoffs, we assume that the core is not stalled by memory.

\textbf{Workloads.}
We choose three workloads from MLPerf to represent different tasks, ResNet50~\cite{kaiming2016resnet} for computer vision, DLRM \cite{maxim19dlrm} for recommendation, and BERT\cite{jacob2019bert} for natural language processing.  
We choose three layers from each workload.
Table~\ref{table:workloads} summarizes the dimensions of the convolution layers in ResNet50 and FC layers in DLRM and BERT that we use. We use the following notation: N for batch size, K for the number of filters, C for the number of input channels, X and Y for input dimensions, R and S for filter dimensions, NIN for the number of input neurons, NON for the number of output neurons. We perform experiments on inference due to the extremely long simulation time for training; however, our proposed concept is not limited to inference since GEMM is also a key building block for training~\cite{Evangelos20ipdps}.

\begin{table}[htbp] 
\caption{Layer Dimensions Used in Evaluation}
\begin{center}
\begin{tabular}{|c|c|}
\hline
\textbf{Layer}&\textbf{Dimensions} \\
\hline
ResNet50-1 & N=32 K=C=64 X=Y=56 R=S=1  \\
ResNet50-2 & N=32 K=C=64 X=Y=56 R=S=3  \\
ResNet50-3 & N=32 K=512 C=1024 X=Y=14 R=S=1  \\
DLRM-1 & N=512 NIN=1024 NON=1024  \\
DLRM-2 & N=512 NIN=1024 NON=64  \\
DLRM-3 & N=512 NIN=2048 NON=2048  \\
BERT-1 & N=256 NIN=768 NON=768  \\
BERT-2 & N=256 NIN=3072 NON=768  \\
BERT-3 & N=256 NIN=768 NON=3072  \\

\hline
\end{tabular}
\label{tab1}
\end{center}
\vspace{-2mm}
\label{table:workloads}
\end{table}
\insertFigure{runtime-comparison.png}{Runtime of different \RASA optimizations normalized to runtime of baseline. The relative performances of various configurations are independent of workloads. 
}

We evaluate the baseline design (no pipelining with baseline PEs) and seven RASA-based designs, whose names indicate
the applied optimizations. For example, RASA-DM-PIPE uses DM and PIPE. For fair comparisons, we use the same number of multipliers in all systolic arrays. We use a $32\times16$ array of PEs ($16\times16$ if DM is applied) to match the tile register dimensions. We run all systolic arrays at 500 MHz.

\textbf{Area Overhead and Energy-Efficiency.}
We implemented and synthesized the \RASA-data optimizations with the Synopsis DC compiler on Nangate 15nm. Then, we used Cadence Innovus for place-and-route to understand area and power costs. 
The area of the baseline systolic array (32$\times$16 baseline PEs) is 0.7\% of the total die size of an Intel Skylake GT2 4C CPU. RASA-DB, RASA-DM, and RASA-DMDB have 3.1\%, 2.6\%, and 5.5\% area overhead over the baseline systolic array, respectively. RASA-DB, RASA-DM, and RASA-DM-DB achieve average energy efficiency improvements vs. the baseline of 4.38$\times$, 2.19$\times$, and 4.59$\times$.

\textbf{Runtime and PPA.} Fig.~\ref{fig:runtime-comparison.png} compares the runtime of each design normalized to the baseline. \RASA-PIPE and \RASA-WLBP reduce runtime by an average of 15.7\% and 30.9\%, respectively.  These designs only require control hardware changes (and eight dirty bits for tile registers with \RASA-WLBP) over the baseline.
\RASA-DB-WLS adds buffers to enable the most aggressive pipelining, achieving a 78.1\% average reduction in runtime.
The remaining designs use DM to merge pairs of PEs.  \RASA-DM-WLBP gives a 55.5\% average improvement in runtime (24.6\% benefit over \RASA-WLBP).  Combining DM and DB, \RASA-DMDB-WLS gives similar performance to \RASA-DB-WLS, a 79.2\% average runtime reduction. 

\insertFigure{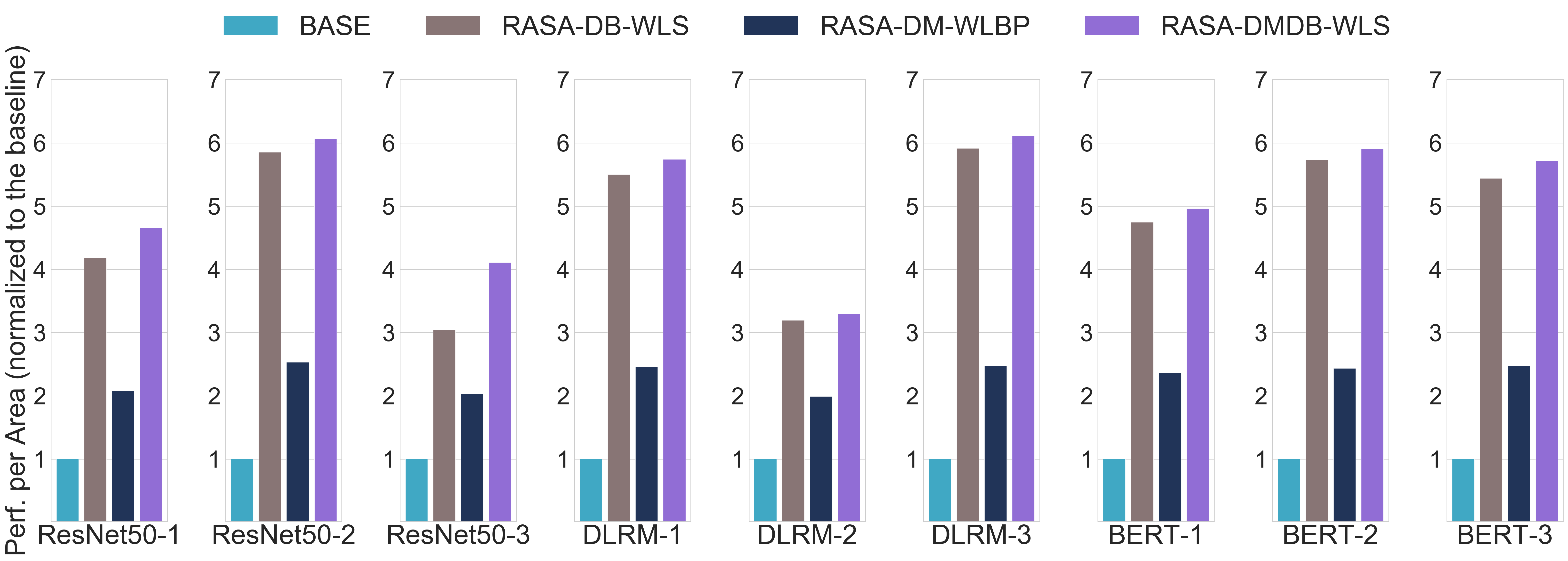}{Performance Per Area (PPA) for different \RASA-Data optimizations.}

In Fig.~\ref{fig:ppa-comparison.png}, we compare three different \RASA-Data optimizations: RASA-DB, RASA-DM, and RASA-DMDB in terms of performance per area (normalized to the baseline). We apply the best-performing \RASA-Control optimization to each \RASA-Data optimization, i.e., RASA-DB-WLS, RASA-DM-WLBP, RASA-DMDB-WLS. Since the area overhead of RASA-Data optimizations are small, performance per area shows the similar trend with runtime.


\insertFigure{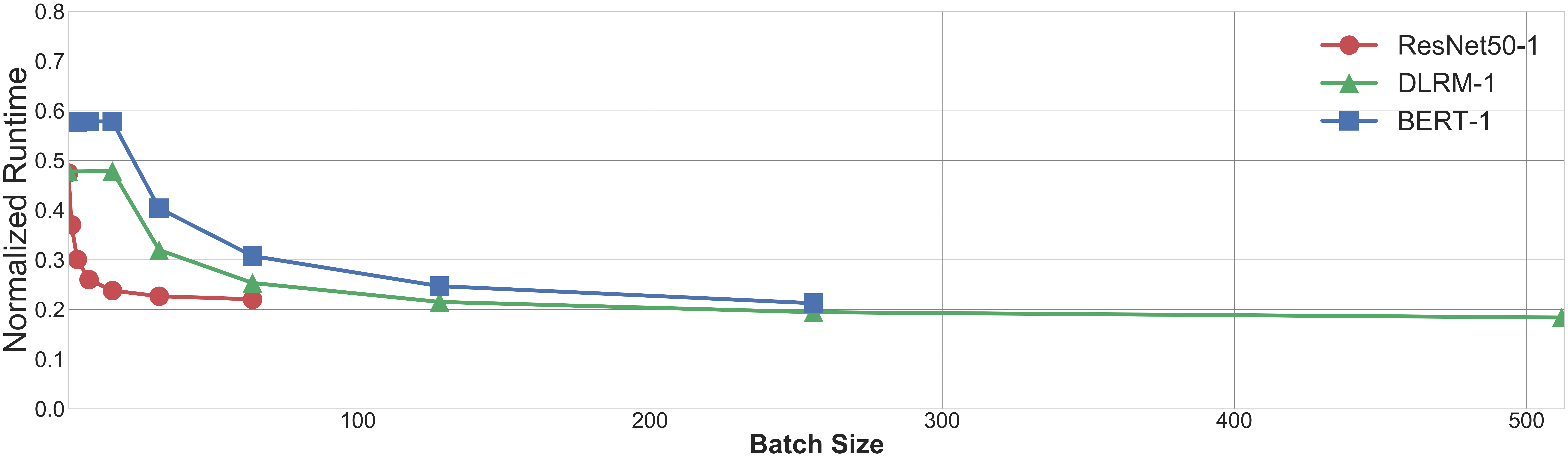}{Normalized runtime of \RASA-DMDB-WLS with different batch sizes.}


\textbf{Sensitivity on Batch Size.}
In Fig.~\ref{fig:sensitivity-batchsize.png}, we explore how \RASA works with different batch sizes. In this graph, we use \RASA-DMDB-WLS to understand the largest performance gain that can be achieved with \RASA optimizations for different batch sizes. 
First, we observe that the FC layers with very small batches (1, 2, 4, 8, and 16), have very similar normalized runtimes due to the systolic array being 16$\times$16 (or 16$\times$32). Thus, these runs all use the same number of \rasamm instructions since 16 is the smallest granularity of work. Second, as batch size increases, runtime approaches an asymptote. For large batches, \rasamm instructions dominate the workload.
${L_{baseline}}=95$ cycles for the configuration in our evaluation (Eq.~\ref{ws_latency}). If we perfectly pipeline all \rasamm, we complete a \rasamm every 16 cycles.  Thus, \RASA-DMDB-WLS can at best bring the normalized runtime down to $\frac{16}{95}=0.168$. 

\section{Conclusions}
\label{sec:conclusions}
This work is the first to study the implications of integrating a matrix engine inside the CPU pipeline.
Specifically, we identify challenges 
with under-utilization of a
systolic array when used as a matrix engine due to frequent fills and drains because of limited sized registers inside CPUs. We propose an efficient pipelined register-aware systolic array
called \RASA that provides 79.2\% runtime improvement and 4.59$\times$ energy efficiency over a baseline systolic array on various DL workloads with negligible area overhead. 

\bibliographystyle{IEEEtran}
\bibliography{references}

\vspace{12pt} 

\end{document}